# CULTURE AND THE DISPOSITION EFFECT

Bastian Breitmayer *

Queensland University of Technology

Tim Hasso * †

Bond University

thasso@bond.edu.au

Matthias Pelster *

University of Paderborn

**Abstract**

We study the relationship between national culture and the disposition effect by investigating international differences in the degree of investors' disposition effect. We utilize brokerage data of 387,993 traders from 83 countries and find great variation in the degree of the disposition effect across the world. We find that the cultural dimensions of long-term orientation and indulgence help to explain why certain nationalities are more prone to the disposition effect. We also find support on an international level for the role of age and gender in explaining the disposition effect.

*All authors contributed equally, author order is the result of randomization

† Corresponding author

# 1. INTRODUCTION

The disposition effect is one of the most studied market anomalies in behavioral finance. It describes the tendency of investors to realize gains while holding onto losing investments (Shefrin & Statman, 1985). As this behavior leads to portfolios that are over-weight on positions that continue to lose value, the disposition effect leads to adverse investor performance and has implications for overall market returns. Studying the determinants of the disposition effect, researchers have reported a variation of the effect across demographic characteristic such as gender and age (Dhar & Zhu, 2006; Rau, 2014). While further studies have documented the disposition effect in various countries, a direct comparison of the effect magnitude and possible reasons for the variation across countries and different cultures is, however, missing (Kaustia, 2011). This is surprising as national culture has been shown to explain the underlying drivers of the disposition effect at the individual level, such as loss aversion (Wang, Rieger, & Hens, 2017) and mental accounting (Banerjee, Chatterjee, Mishra, & Mishra, 2019). Furthermore, the meta-analysis by Taras, Kirkman, and Steel (2010) showed that Hofstede's cultural dimensions can predict emotions at an individual level. Because emotions may explain the disposition effect (Kaustia, 2011), it stands to reason that people from different cultures vary in their disposition effect because of differences in emotion regulation.

In this paper we address this gap in research by studying the variation in the disposition effect across 83 countries using individual investors' brokerage data, and explore whether cultural factors may help to explain this variation. Using international data from a single broker allows us to move outside the laboratory into the real world, while still ensuring that investors use the same trading platform to maximize internal validity.

We find that there is substantial international variation in the degree of the disposition effect, with the national averages ranging from -0.04 to 0.22. We document that the cultural dimensions of long-term orientation as well as indulgence are related to decreased levels of disposition effect. Furthermore, our results are consistent with prior findings suggesting that the disposition effect is greater for females and older people. Consequently, we establish that these patterns hold on an international level.

## 2. DATA

We use transaction-level brokerage data from a UK-based broker. While the broker operates under a UK license its clientele is international as clients open accounts online. The broker allows clients to trade contracts-for-difference (CFDs) on a variety of instruments including stocks, commodities, and currency pairs among others. In total we have full trading histories of 668,067 investors from 1$^{st}$ of January 2014 to 31$^{st}$ of December 2017. However, as we have incomplete demographical information for some investors and others are from countries where Hofstede's cultural dimensions have not been measured, we exclude these from our formal testing. As such, our final sample consists of 387,993 traders from 83 countries. We define nationality based on the citizenship of the investor, which is verified by the broker. The number of investors per country are of unequal size, with the ten largest countries in our sample being: (1) United Kingdom (n=60,226); (2) Germany (n=49,319); (3) France (n=28,771); (4) Italy (n=24,569); (5) Brazil (n=14,958); (6) Spain (n=14,797); (7) Morocco (n=13,918); (8) Canada (n=11,032); (9) Russia (n=9,823); (10) Australia (n=8,669).

We use Hofstede's six cultural dimensions to capture the country culture (Hofstede, 2001). The six dimensions are power distance, individualism, masculinity, uncertainty avoidance, long-term orientation, and indulgence. We collect the culture data from Hofstede Insights. We also collect country economic conditions data from the World Bank, which includes the GDP per capita as well as the GDP per capita growth.

## 3. METHOD

We estimate the disposition effect of each investor in our sample by following the approach of Odean (1998). For each trading day, we count the number of realized gains and losses, as well as the number of paper gains and losses for each investor. We then use this information to estimate the disposition effect for each investor using the following equation:

$$Disposition\ Effect = \frac{realized\ gains}{realized\ gains + paper\ gains} - \frac{realized\ losses}{realized\ losses + paper\ losses} \quad (1)$$

The variable has a theoretical range of -1 to +1, where -1 represents an investor that holds all positions that are in-the-money and sells all positions that are out-of-the-money, +1 represents an investor that sells are positions that are in-the-money and holds all positions that are out-of-the-money, and 0 represents an investors that equally sells and holds in-the-money and out-of-the-money positions. Consequently, if the variable value is above 0, an investor displays the disposition effect, with higher values indicating a greater effect.

To understand what explains the variation in the disposition effect we then estimate the following regression:

$$Disposition\ Effect_i = a + b\ Culture_i + c\ Demographics_i + d\ Economic_i + e\ Region_i + \varepsilon_i$$
$$(2)$$

where disposition effect is defined as per equation 1 and calculated on an average basis for each investor in the sample. Culture is a vector of six culture variables using Hofstede's dimensions, each ranging from 0 to 100. Demographics is a vector of gender and age dummy variables where Male is a dummy coded as 1 if the investor is male. Age are dummy variables coded according to the age brackets of 18-24, 25-34, 35-44, 45-54, 55-64, and >65. Economic is a vector of continuous economic condition variables that includes GDP per capita (log) and GDP per capital growth. Region is a vector of dummy coded variables capturing the region of the investor. We estimate the regression using ordinary least squared, using robust standard errors that are clustered based on nationality.

## 4. RESULTS

We use Figure 1 to visually report the average disposition effect for each country in our sample. The average disposition effect ranges from -0.04 to 0.22. The variation in disposition effects provides the justification for further analysis into understanding why differences exist on an international basis.

We report our primary results in Table 1. We estimate five models by considering possible explanatory variable categories, where model five is the full model including all variables. We focus our discussion on the results of model five. Similar to previous studies (Rau, 2014), we find that men suffer from the disposition effect to a lesser extent, and that the disposition effect appears to increase with age. We do not find that country economic conditions have an effect after we control for culture. We find that investors in countries in the Asia-Pacific present higher disposition effects as compared to Europeans and investors in the Sub-Saharan Africa. We find that two of Hofstede's cultural dimensions help to explain the variation in the disposition effect. Both long-term orientation and indulgence have significant negative relationships with the disposition effect. Economically, the impact of the cultural variables on the disposition is quite meaningful. In particular, at the mean a one standard deviation increase in long-term orientation and indulgence is associated with a 15.7% and 25.2% decrease of the disposition effect (mean = 0.024), respectively. In Figure 2, we graph the linear predictions for the disposition effect based on long-term orientation and indulgence.

As some countries in our sample only have a small number of investors we perform robustness tests to ensure that this does not drive our results. We compare our main model to restricted samples using minimum investor thresholds (n>100, n>500, and n>1000). In each stage of our robustness, the significance, direction, and magnitude are stable across the models (untabulated).

## 5. DISCUSSION

In this paper we explored the relationship between national culture and the disposition effect. As such, we addressed a gap in previous research around the disposition effect where previous work has been dominated by single country studies, with the majority focusing on US investors (Kaustia, 2011). We found that the cultural dimensions of long-term orientation and indulgence are negatively related with the disposition effect. Long-term orientation refers to the country's ability to focus on the future in a pragmatic way and accepting changing conditions by for example investing in education instead of clinging to the past (Hofstede, 2001). This finding can be related to previous work on how education has the ability to decrease the disposition effect. Indulgence on the other hand refers to the country's acceptance of free gratification, having fun, and the rejection of strict social norms (Hofstede, 2001). This finding is less easily explained based on past work but can perhaps be explored by considering the association between conservative societies and loss aversion, as loss aversion is one of the key driving elements behind the disposition effect (Shefrin & Statman, 1985).

Our results suffer from limitations around our sample. As we use brokerage data our study suffers from the limitation of self-selection. However, we believe that it would be incredibly difficult to provide representative international data on an international scale using real trading behavior, and even experimental work would be immensely difficult given the magnitude of data collection that would have to take place for representative findings (which may still suffer from external validity). As such, we provide a valuable contribution to the literature and our results provide an avenue for future work on this area to untangle the relationship between culture and the disposition effect.

**Table 1. Explaining international variation in disposition effects**

|  | (1) | (2) | (3) | (4) | (5) |
|---|---|---|---|---|---|
| *Demographic - Gender* | | | | | |
| Male | -0.0149*** | | | | -0.0142*** |
|  | (0.00270) | | | | (0.00266) |
| *Demographic - Age* | | | | | |
| 25-34 | 0.00853*** | | | | 0.00846*** |
|  | (0.00154) | | | | (0.00135) |
| 35-44 | 0.0141*** | | | | 0.0157*** |
|  | (0.00254) | | | | (0.00154) |
| 45-54 | 0.0128*** | | | | 0.0164*** |
|  | (0.00298) | | | | (0.00157) |
| 55-64 | 0.0107*** | | | | 0.0143*** |
|  | (0.00344) | | | | (0.00297) |
| >65 | 0.0224*** | | | | 0.0247*** |
|  | (0.00502) | | | | (0.00516) |
| *Economic* | | | | | |
| GDP per capita (log) | | -0.0109*** | | | -0.00483 |
|  | | (0.00217) | | | (0.00375) |
| GDP per capita growth | | 0.00162 | | | 0.000354 |
|  | | (0.000978) | | | (0.000973) |
| *Region* | | | | | |
| Europe | | | -0.0174** | | -0.0100* |
|  | | | (0.00742) | | (0.00586) |
| Latin America-Caribbean | | | -0.00620 | | -0.00595 |
|  | | | (0.00693) | | (0.00678) |
| Middle East-North Africa | | | 0.0188** | | 0.00202 |
|  | | | (0.00837) | | (0.00917) |
| North America | | | -0.0160** | | -0.00316 |
|  | | | (0.00682) | | (0.00621) |
| Sub-Saharan Africa | | | -0.0243*** | | -0.0264*** |
|  | | | (0.00732) | | (0.00901) |
| *Culture* | | | | | |
| Power Distance | | | | 0.000116 | 3.57e-05 |
|  | | | | (7.68e-05) | (8.67e-05) |
| Individualism | | | | -0.000170** | -2.25e-05 |
|  | | | | (6.67e-05) | (8.30e-05) |
| Masculinity | | | | 8.78e-05 | 2.58e-05 |
|  | | | | (7.22e-05) | (6.67e-05) |
| Uncertainty avoidance | | | | 1.02e-05 | 0.000111 |
|  | | | | (6.33e-05) | (6.99e-05) |
| Long-term orientation | | | | -0.000326*** | -0.000179** |
|  | | | | (6.60e-05) | (8.15e-05) |
| Indulgence | | | | -0.000470*** | -0.000309*** |
|  | | | | (9.03e-05) | (0.000115) |
| Constant | 0.0279*** | 0.131*** | 0.0364*** | 0.0641*** | 0.0986** |
|  | (0.00364) | (0.0225) | (0.00681) | (0.0136) | (0.0433) |
|  | | | | | |
| Observations | 387,993 | 387,993 | 387,993 | 387,993 | 387,993 |
| R-squared | 0.1% | 0.2% | 0.2% | 0.3% | 0.4% |

Note. OLS regressions where the dependent variable is disposition effect. Robust standard errors (in parentheses) are adjusted for 83 clusters in nationality. Female, 18-24, and Asia-Pacific are baselines for their respective groups. *** p<0.01, ** p<0.05, * p<0.1

**Figure 1. Average Disposition Effect by Country**

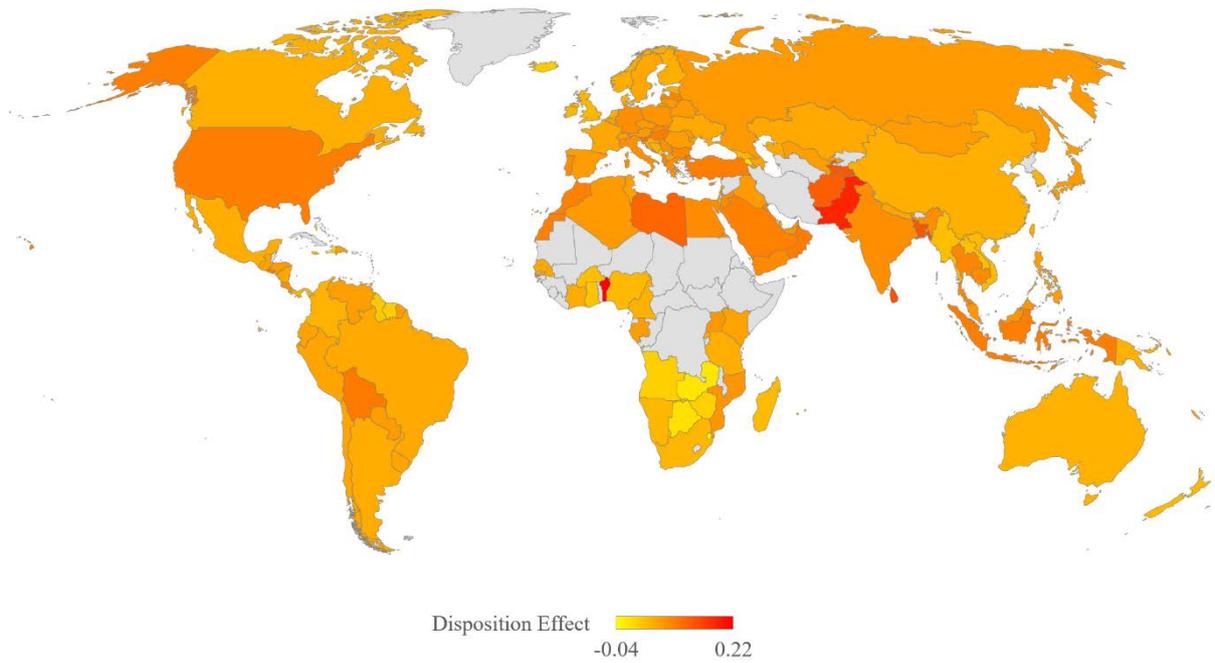

*Note. Figure illustrates average disposition effect by country. The figure includes countries with at least one hundred unique investors.*

**Figure 2. Linear Prediction of Disposition Effect**

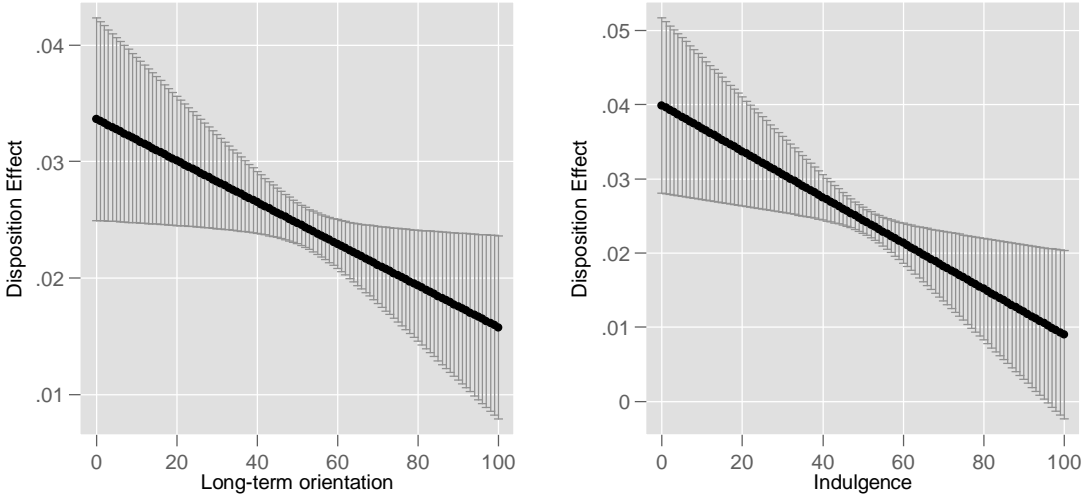

*Note. Figure illustrates the linear prediction and 95% confidence interval of the disposition effect based on assumed intervals of 0 to 100 for two of Hofstede's cultural dimensions. The linear prediction is based on the results of an OLS regression as specified in Model 5 of Table 1.*